\documentclass{article}

\usepackage{PRIMEarxiv}

\usepackage[utf8]{inputenc} 
\usepackage[T1]{fontenc}    
\usepackage{hyperref}       
\usepackage{url}            
\usepackage{booktabs}       
\usepackage{amsfonts}       
\usepackage{nicefrac}       
\usepackage{microtype}      
\usepackage{lipsum}
\usepackage{fancyhdr}       
\usepackage{graphicx}       
\usepackage{amsmath,amssymb}

\newcommand{\T}{^{^{_T}}}

\title{A Collaborative, Interactive and Context-Aware Drawing Agent for Co-Creative Design
}

\author{
  Francisco Ibarrola\\
  The University of Sydney,\\
  NSW, Australia.\\
  \scalebox{.9}[1.0]{\texttt{francisco.ibarrola@sydney.edu.au}}\\
  \And
  Tomas Lawton\\
  The University of Sydney,\\
  NSW, Australia.\\
  \scalebox{.9}[1.0]{\texttt{tlaw8603@uni.sydney.edu.au}}\\
  \And
  Kazjon Grace \\
  The University of Sydney,\\
  NSW, Australia.\\
  \scalebox{.9}[1.0]{\texttt{kazjon.grace@sydney.edu.au}} \\
}

\begin{document}
\maketitle

\begin{abstract}
Recent advances in text-conditioned generative models have provided us with neural networks capable of creating images of astonishing quality, be they realistic, abstract, or even creative. These models have in common that (more or less explicitly) they all aim to produce a high-quality one-off output given certain conditions, and in that they are not well suited for a creative collaboration framework. Drawing on theories from cognitive science that model how professional designers and artists think, we argue how this setting differs from the former and introduce CICADA: a Collaborative, Interactive Context-Aware Drawing Agent. CICADA uses a vector-based synthesis-by-optimisation method to take a partial sketch (such as might be provided by a user) and develop it towards a goal by adding and/or sensibly modifying traces. Given that this topic has been scarcely explored, we also introduce a way to evaluate desired characteristics of a model in this context by means of proposing a diversity measure. CICADA is shown to produce sketches of quality comparable to a human user’s, enhanced diversity and most importantly to be able to cope with change by continuing the sketch minding the user’s contributions in a flexible manner.
\end{abstract}

\keywords{Human-Computer Interaction \and Computational Creativity \and Image Generation}

\section{Introduction}

The most common conception of a computer within the process of drawing is that of something enabling and unobtrusive: a tool. While that mindset is in accordance with the actual computer capabilities of up to a few years ago, recent advances (particularly generative models in the fields of imaging and language processing \cite{frans2021clipdraw, ramesh2022hierarchical}) offer a potential reframing, permitting enough for us to start looking at computers as co-creators \cite{davis2013human}.

This co-creative, agentive role for creative AI offers an opportunity to address one of the biggest shortcomings of generative AI to date. In the creative professions, from art to interface design to composition to architecture, the primary purpose of early sketching is conceptual exploration.  Designers not only tend not to know what they are looking for when they start a creative process \cite{gero1998conceptual}, but those that think they do tend to fail \cite{dorst2015frame}. Designing is not --at least in its early stages-- a process of synthesis given some requirements, but instead a co-evolution of problem and solution \cite{poon1997co}: an iterative process of discovering and refining requirements in conjunction with design behaviour and structure \cite{gero2004situated}. If generative AI is to seriously augment human creativity, these ideas from design cognition need to move from psychology journals into our cost functions and model architectures.

From the perspective of building a co-creative design tool, this means that is not sufficient to focus on the problem of generating useful, novel or diverse proposals. Instead we need to see the task as a collaborative user/system process characterised by both sides developing, refining, and aligning their intentions and adapting to each others' changing goals. By contrast, most existing applications of generative AI to creative domains focus on generating outputs of high visual quality, typically conditioned on a specific stimulus. In this paper we discuss how generative models based on multimodal contrastive training could be designed to be effective aids to human--machine collaborative creativity.

In light of this, our current work shall focus on three aspects. Firstly, we must introduce this somewhat underexplored problem (at least in the Machine Learning (ML) literature) and define the key requirements that a successful co-creative agent must address. Secondly, we build such an agent. And third, given the novelty of the co-creative AI problem, we propose (and apply) some ways to test our agent against the aforementioned requirements. We further argue the robustness of our approach to additional technological advances in this field.

\setlength{\fboxrule}{1.2pt}%
\begin{figure*}[h]
    \centering
    
    \begin{minipage}{0.095\textwidth}
    \small
    \centering
    A Red Chair
    \end{minipage}
    \hspace{0.3cm}
    \begin{minipage}{0.095\textwidth}
    \small
    \centering
    A Drawing of a Hat
    \end{minipage}
    \hspace{0.3cm}
    \begin{minipage}{0.095\textwidth}
    \small
    \centering
    A Drawing of a Lamp
    \end{minipage}
    \hspace{0.3cm}
    \begin{minipage}{0.095\textwidth}
    \small
    \centering
    A Drawing of a Pot
    \end{minipage}
    \hspace{0.3cm}
    \begin{minipage}{0.095\textwidth}
    \small
    \centering
    A Drawing of a Boat
    \end{minipage}
    \hspace{0.3cm}
    \begin{minipage}{0.095\textwidth}
    \small
    \centering
    A Blue Dress
    \end{minipage}
    \hspace{0.3cm}
    \begin{minipage}{0.095\textwidth}
    \small
    \centering
    A High-Heel Shoe
    \end{minipage}
    \hspace{0.3cm}
    \begin{minipage}{0.095\textwidth}
    \small
    \centering
    A Drawing of a Bust
    \end{minipage}
    
    \vspace{0.1cm}
    
    \fbox{\includegraphics[width=0.095\textwidth]{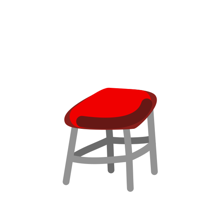}}
    \hspace{0.01cm}
    \fbox{\includegraphics[width=0.095\textwidth]{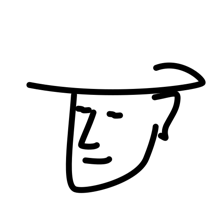}}
    \hspace{0.01cm}
    \fbox{\includegraphics[width=0.095\textwidth]{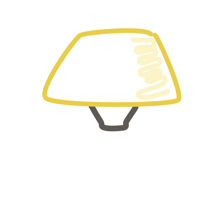}}
    \hspace{0.01cm}
    \fbox{\includegraphics[width=0.095\textwidth]{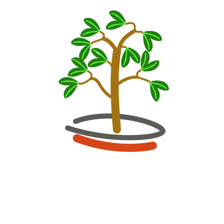}}
    \hspace{0.01cm}
    \fbox{\includegraphics[width=0.095\textwidth]{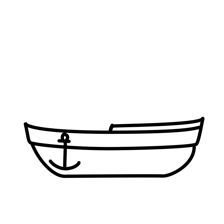}}
    \hspace{0.01cm}
    \fbox{\includegraphics[width=0.095\textwidth]{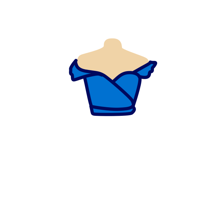}}
    \hspace{0.01cm}
    \fbox{\includegraphics[width=0.095\textwidth]{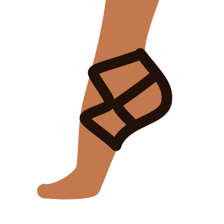}}
    \hspace{0.01cm}
    \fbox{\includegraphics[width=0.095\textwidth]{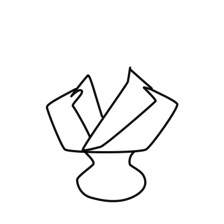}}
    
    \vspace{0.5cm}
    
    \fbox{\includegraphics[width=0.095\textwidth]{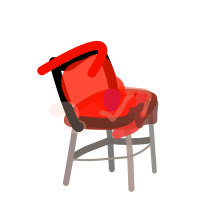}}
    \hspace{0.01cm}
    \fbox{\includegraphics[width=0.095\textwidth]{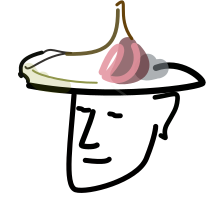}}
    \hspace{0.01cm}
    \fbox{\includegraphics[width=0.095\textwidth]{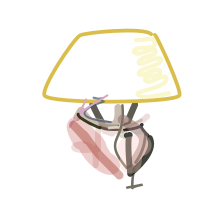}}
    \hspace{0.01cm}
    \fbox{\includegraphics[width=0.095\textwidth]{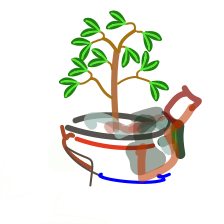}}
    \hspace{0.01cm}
    \fbox{\includegraphics[width=0.095\textwidth]{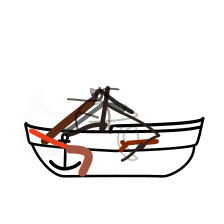}}
    \hspace{0.01cm}
    \fbox{\includegraphics[width=0.095\textwidth]{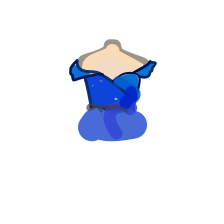}}
    \hspace{0.01cm}
    \fbox{\includegraphics[width=0.095\textwidth]{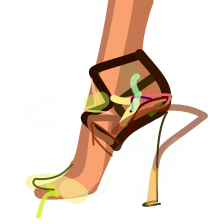}}
    \hspace{0.01cm}
    \fbox{\includegraphics[width=0.095\textwidth]{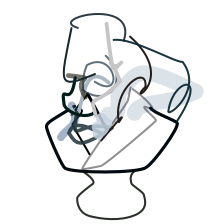}}
    
    \vspace{0.1cm}
    
    \fbox{\includegraphics[width=0.095\textwidth]{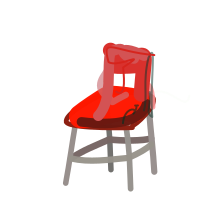}}
    \hspace{0.01cm}
    \fbox{\includegraphics[width=0.095\textwidth]{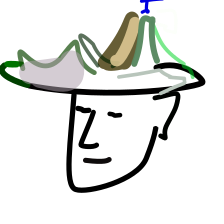}}
    \hspace{0.01cm}
    \fbox{\includegraphics[width=0.095\textwidth]{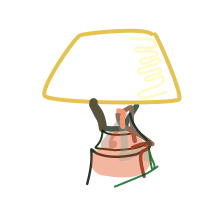}}
    \hspace{0.01cm}
    \fbox{\includegraphics[width=0.095\textwidth]{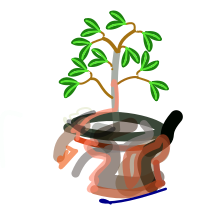}}
    \hspace{0.01cm}
    \fbox{\includegraphics[width=0.095\textwidth]{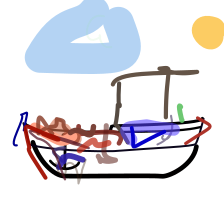}}
    \hspace{0.01cm}
    \fbox{\includegraphics[width=0.095\textwidth]{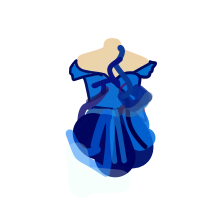}}
    \hspace{0.01cm}
    \fbox{\includegraphics[width=0.095\textwidth]{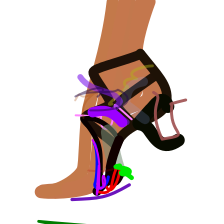}}
    \hspace{0.01cm}
    \fbox{\includegraphics[width=0.095\textwidth]{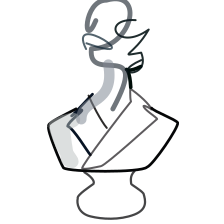}}
    
    \vspace{0.1cm}
    
    \fbox{\includegraphics[width=0.095\textwidth]{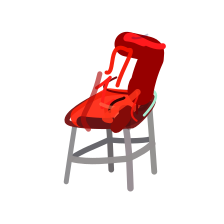}}
    \hspace{0.01cm}
    \fbox{\includegraphics[width=0.095\textwidth]{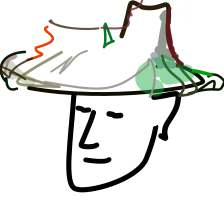}}
    \hspace{0.01cm}
    \fbox{\includegraphics[width=0.095\textwidth]{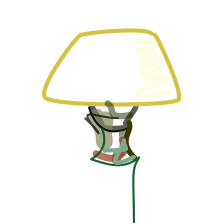}}
    \hspace{0.01cm}
    \fbox{\includegraphics[width=0.095\textwidth]{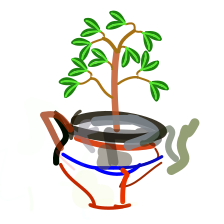}}
    \hspace{0.01cm}
    \fbox{\includegraphics[width=0.095\textwidth]{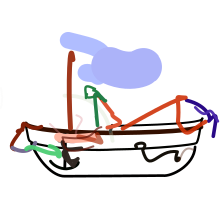}}
    \hspace{0.01cm}
    \fbox{\includegraphics[width=0.095\textwidth]{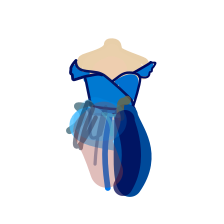}}
    \hspace{0.01cm}
    \fbox{\includegraphics[width=0.095\textwidth]{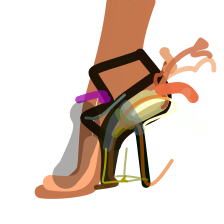}}
    \hspace{0.01cm}
    \fbox{\includegraphics[width=0.095\textwidth]{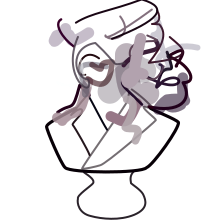}}
    
    \vspace{0.1cm}
    
    \fbox{\includegraphics[width=0.095\textwidth]{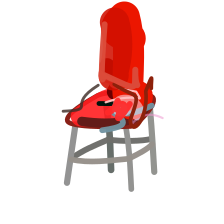}}
    \hspace{0.01cm}
    \fbox{\includegraphics[width=0.095\textwidth]{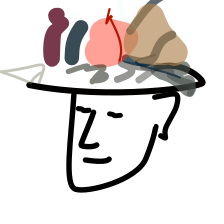}}
    \hspace{0.01cm}
    \fbox{\includegraphics[width=0.095\textwidth]{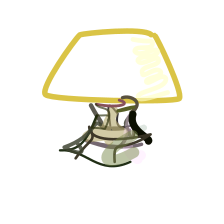}}
    \hspace{0.01cm}
    \fbox{\includegraphics[width=0.095\textwidth]{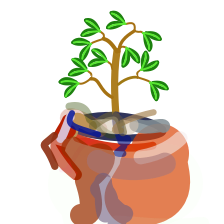}}
    \hspace{0.01cm}
    \fbox{\includegraphics[width=0.095\textwidth]{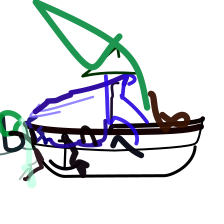}}
    \hspace{0.01cm}
    \fbox{\includegraphics[width=0.095\textwidth]{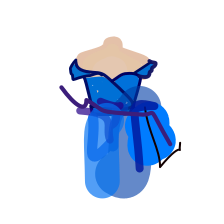}}
    \hspace{0.01cm}
    \fbox{\includegraphics[width=0.095\textwidth]{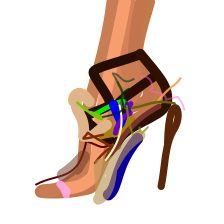}}
    \hspace{0.01cm}
    \fbox{\includegraphics[width=0.095\textwidth]{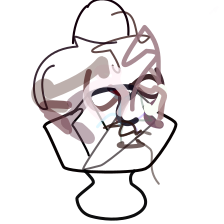}}
    
    \vspace{0.1cm}
    
    \fbox{\includegraphics[width=0.095\textwidth]{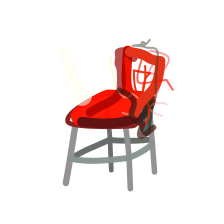}}
    \hspace{0.01cm}
    \fbox{\includegraphics[width=0.095\textwidth]{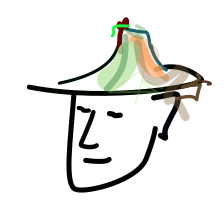}}
    \hspace{0.01cm}
    \fbox{\includegraphics[width=0.095\textwidth]{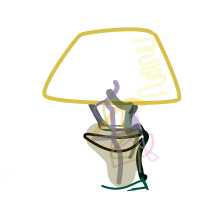}}
    \hspace{0.01cm}
    \fbox{\includegraphics[width=0.095\textwidth]{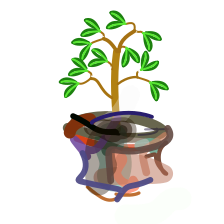}}
    \hspace{0.01cm}
    \fbox{\includegraphics[width=0.095\textwidth]{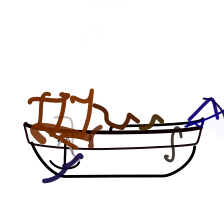}}
    \hspace{0.01cm}
    \fbox{\includegraphics[width=0.095\textwidth]{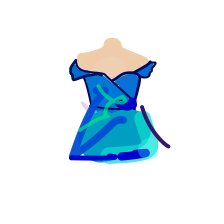}}
    \hspace{0.01cm}
    \fbox{\includegraphics[width=0.095\textwidth]{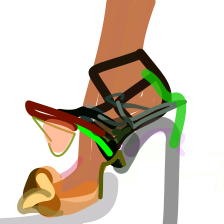}}
    \hspace{0.01cm}
    \fbox{\includegraphics[width=0.095\textwidth]{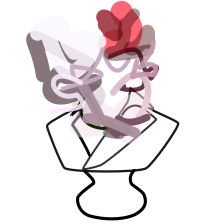}}
    
    \vspace{0.1cm}
    
    \fbox{\includegraphics[width=0.095\textwidth]{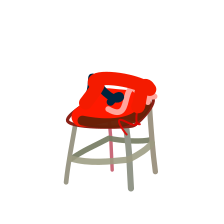}}
    \hspace{0.01cm}
    \fbox{\includegraphics[width=0.095\textwidth]{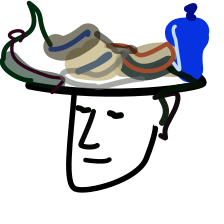}}
    \hspace{0.01cm}
    \fbox{\includegraphics[width=0.095\textwidth]{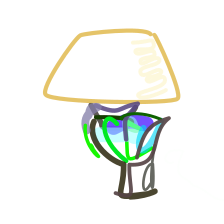}}
    \hspace{0.01cm}
    \fbox{\includegraphics[width=0.095\textwidth]{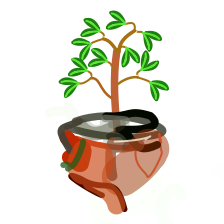}}
    \hspace{0.01cm}
    \fbox{\includegraphics[width=0.095\textwidth]{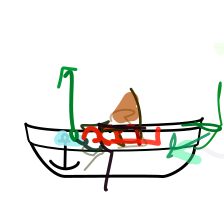}}
    \hspace{0.01cm}
    \fbox{\includegraphics[width=0.095\textwidth]{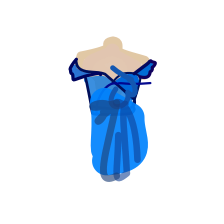}}
    \hspace{0.01cm}
    \fbox{\includegraphics[width=0.095\textwidth]{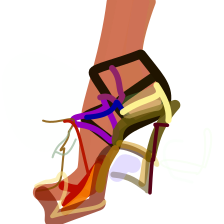}}
    \hspace{0.01cm}
    \fbox{\includegraphics[width=0.095\textwidth]{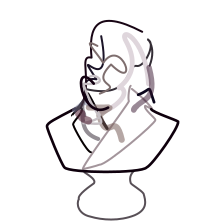}}

    \caption{Results obtained using CICADA for drawing completion (after background pruning). The used prompt is on top of each column, along with the partial sketch immediately below. The subsequent rows depict results obtained with different random initialisations.}
    \label{fig:cicada_gen}
\end{figure*}

\subsection{Operationalising Co-creative Drawing Systems}

Designers sketch to think.  The vast majority of sketches are not intended to be part of a complete design, but instead serve to further the designer's understanding of the problem. Creative design is, by definition, the solving of underspecified problems \cite{gero1990design}, and thus developing the required level of specification is a key component of early portions of the design process. This has been described as a parallel co-evolution of problem and solution \cite{poon1997co}, as it's impossible to properly search the space of possible solutions to an ill-defined problem without also searching the space of full definitions of that problem. We are developing generative models capable of augmenting human performance during this uniquely fluid cognitive process. Existing computational models of this creative design have developed the notion of ``creative meta-search'' \cite{boden2004creative,wiggins2006preliminary}: the exploration of a problem-space that serves to define the subsequent solution-space. Changes in problem-space have also been suggested as a way to evaluate whether a solution is in fact creative -- ``transformational creativity'', solutions that exist outside of the solution space as originally formulated \cite{ritchie2006transformational}. 

From the perspective of the designer, this process has been modelled as a kind of iterative explore/exploit trade-off, a ``reflective conversation with the materials of the situation'' \cite{schon1992designing}. To bring the discussion back to the AI-assisted drawing setting: designers draw, reflect on what they see in their drawings, and in doing so potentially refine their approach to the problem. Designers who engage in this process of acting, reflecting, and revising their approach more frequently have been shown to produce more creative output \cite{suwa2000unexpected}. Our purpose in raising all of this literature from cognitive science is to show that the overwhelming weight of evidence suggests that for generative models to effectively work alongside designers in the early phases of design, they must be focused on helping shape problems, not produce solutions.

To begin operationalising these ideas, consider a blank canvas on which a user intends to sketch out some ideas for a design, given an initial set of goals, requirements, or description. The output is not required to be a finished design solution, nor is there any mention of a specific design domain. The intent is to make progress on some artefact, which might be anything -- a magazine cover, a house, an outfit, a website, a machine part, or a corporate strategy, to name a few. Making progress on the design does not (solely) mean drawing the sketch, it means iteratively exploring and refining ideas. We call this task ``conceptual designing'' \cite{gero1998conceptual}, or more colloquially: ``sketching''. Our goal is to build an ML system that can, in real-time, augment a human's performance on this task, supporting productive and creative human-machine collaboration .

In this paper, we present this framing of human-machine collaborative conceptual designing as a new focus for generative ML (and more specifically text-to-image models). As a user draws, the kind of model we describe should contribute meaningfully in real-time, not only to the lines being drawn but also to the emerging design concept. These contributions need to be contextualised both to the current state of the artefact (in our case, the sketch) and the manifest intentions of the user (in our case, the prompt). To do this we have developed CICADA, a Collaborative, Interactive and Context Aware Drawing Agent that can cope with these requirements.

At any point in time, CICADA's goal is to progress the user's partial sketch in a creative manner, complying with provided intent (in the form of text prompts). ``Creative'' is a strongly contested term \cite{grace2015data}, but perhaps the most well-established definition (at least of a creative artefact or idea) is ``something that is both novel/surprising and valuable/useful'' \cite{amabile2018creativity}. In the context of collaborative drawing we adopt the operationalisation of these two factors used in \cite{mccormack2022quality}: \emph{quality} (utility measured with respect to some domain-dependent metrics) and novelty proxied as \emph{diversity} (the novelty of a set of artefacts measured as their coverage of a particular space) (\cite{mccormack2022quality}).  However, we do not want to focus solely on the outcomes of the sketching process, both because our task is not to complete the design, and our setting features a human in the loop, making it difficult to determine who contributed what.  We shall hence add \emph{awareness} (the degree to which the system respects the existing partial sketch) to the above two qualities. We say awareness rather than fidelity, because we are not referring to keeping the user's traces intact, but to respect them conceptually, making sensible changes where appropriate, as creative tasks often entail.

CICADA must also be designed with conceptual pivots originating from the user in mind.  This means that it should be able to cope with significant changes to both the sketch and prompt(s) without throwing everything away and starting over. CICADA should, in summary, be able to produce high quality and diverse outputs even after significant and iterative changes to both the prompt and the partial sketch throughout the process. 

In the rest of this paper we define a CICADA model and demonstrate how -- unlike preexisting methods -- it aligns with the aforementioned requirements simultaneously, and is thus suitable for co-creative design tasks.

\begin{figure*}[h]
    \centering
    \includegraphics[width=\textwidth]{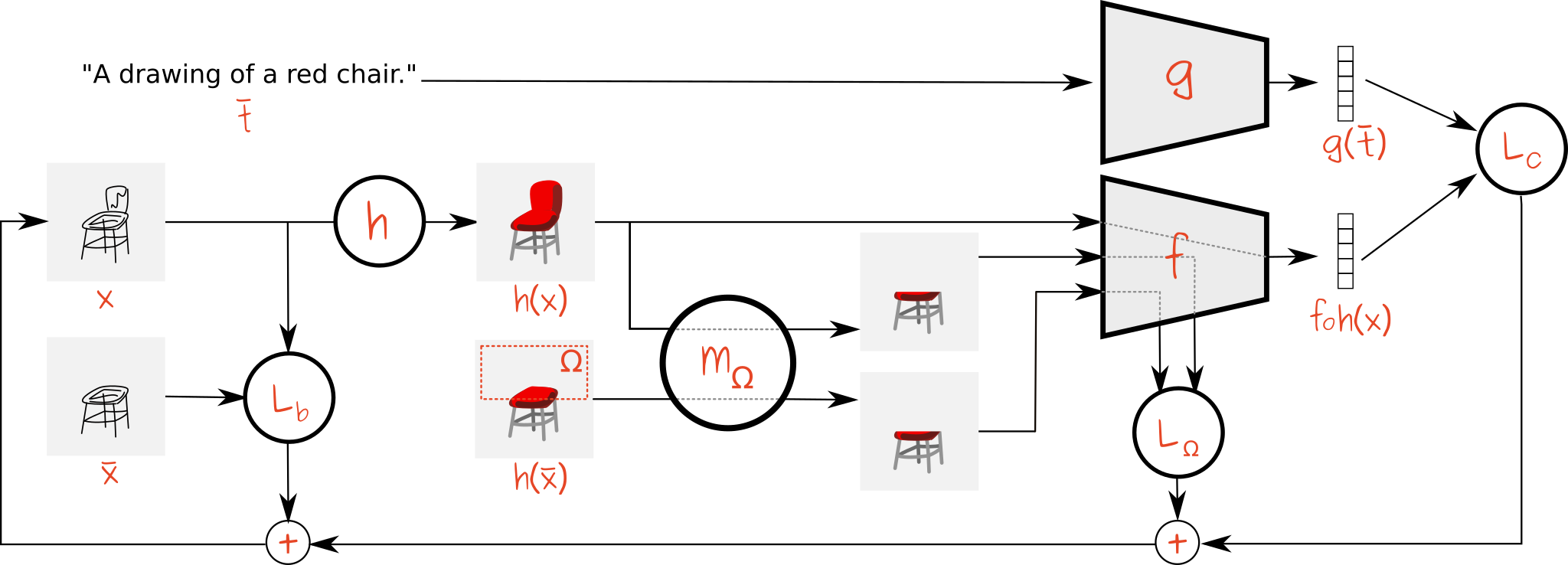}
    \caption{CICADA's drawing model scheme.}
    \label{fig:model_scheme}
\end{figure*}

\subsection{Related Work}

While a diverse array of AI-based drawing systems has been introduced over the last few decades \cite{cohen1995further,colton2012painting,davis2013human,karimi2018computational,lin2020your}, there are two types most relevant to our work. Firstly, those conditional generative models that encompass an interactive component (though they primarily work with raster images) and secondly, those that work with vector graphics, even when most of them are designed for fully autonomous operation.

\subsubsection{Interactive, conditional generative approaches}

As  deep generative networks such as GANs, VAEs and Difussion Models became easier to condition on input stimuli, they began to be used in interactive, human-guided settings. Algorithms like pix2pix \cite{isola2017image} and GauGAN \cite{park2019gaugan} generate natural images from segmentation maps, which can be drawn by users in real time using a painting interface.
Hand-sketching and painting interfaces have also been used for editing images' characteristics, such as of facial features \cite{olszewski2020intuitive}. With the advent of CLIP \cite{radford2021learning} and other models that match images and text together, interactive generators are being developed that leverage both text prompts and the provision of freehand masks \cite{nichol2021glide,ramesh2022hierarchical,avrahami2022blended}.  These semantic inpainting systems, like earlier interactive conditional models, have the primary goal of synthesising a raster image based on user input -- they do not suit our conceptual design setting.

\subsubsection{Vector-based generative approaches}

In order to enable a user and our agent to interact directly over the same sketch, we sought to base our model in vector images. An earlier example of this approach is Sketch-RNN, in which an autoregressive VAE (with LSTM encoder and decoder) was trained to generate class-conditioned sketches \cite{ha2017neural}. Training was made possible by the large-scale QuickDraw dataset, which contains about 75,000 sketches in each of several hundred classes. A major landmark in vector-based generation, Sketch-RNN could auto-complete, interpolate between, and refine drawings. Later work introduced adversarial training, significantly increasing the quality of generated sketches \cite{lu2019sketch}. The significant limit of these systems is data availability: recurrent VAEs require very large amounts of training data, and the QuickDraw dataset, while certainly very large, is limited to novice sketches in several hundred classes.

The model in \cite{ibarrola2022towards}, on which this work is based, is itself based on the synthesis-by-optimisation approach in CLIPDraw (\cite{frans2021clipdraw}). CLIPDraw combines a differentiable rasteriser (\cite{li2020differentiable}) with Contrastive Language-Image Pretraining (CLIP, \cite{ramesh2021zero}) to produce images that match a certain description $\bar{t}$, by minimising the cost function
\begin{align} \label{eqn:semantic_loss}
    L_s(x; \bar{t}) \doteq  -\langle g(\bar{t}), f \circ h(x)\rangle
\end{align}
with respect to $x$, where $\langle \cdot , \cdot \rangle$ denotes the cosine similarity, $h$ is the differentiable rasteriser and $g$ and $f$ are Neural Networks encoding the text prompt $t$ and the rasterised image into a latent space. We adapt this by incorporating a partial geometric loss function that weights in the image resemblance to the user's partial sketch and an initialisation and a pruning method to modify CLIPDraw's characteristic outputs into more sketch-completion suitable ones. Given that curve-fitting is a gradient descent process and the initialisation gives a compelling start point, optimisation can occur in real time as the user draws, resulting in an overall more fluid and interactive experience. The model is explained in more detail in the next section.

\section{The CICADA method}

In what follows we will work under the assumption that the partial sketch provided by the user is in the form of a set of $K_0$ B\'ezier curves $\bar{x}$, where $\bar{x}_k^{(1)} \in \mathbb{R}^{D_k\times 2}$ is a set of coordinates, $\bar{x}_k^{(2)} \in[0,1]^4$ is the RGBA stroke colour vector, and $\bar{x}_k^{(3)}>0$ is the stroke width -- the width of each curve as it is traced. These hypotheses reflect the industry standard representation of simple un-filled vector paths as encoded in formats like SVG. This framework is advantageous by comparison to methods that work directly with raster images because of the dramatically smaller number of parameters available to the user: they can select any subset of the points or properties of the B\'ezier curves and modify them throughout the design process. Furthermore, both user-drawn and system-drawn curves can be modified in the same way.

Additionally, it is useful in some instances to request that the user provide a broadly defined drawing region $\Omega$ as a subset of the canvas. This allows the user to specify what part of the existing sketch the agent should modify or refrain from modifying -- it is intended to be used as a suggested drawing space rather than providing a hard constrain as would be expected in an inpainting task, even if it could function like that given sufficiently large penalization. Additionally, this helps by providing cues for a good initialisation of the curves.

\subsection{Curve Fidelity}

We want to enforce curve fidelity -- by which we mean restrict the degree to which the system changes the curves drawn by the user -- so that the model's output is still recognisably consistent with the user's drawing. One way of enforcing this, used in \cite{ibarrola2022towards}, is to penalise the discrepancies between $\bar{x}$ and the first $K_0$ curves our agent must draw. To do that, let us denote by $K_a$ the number of additional curves being drawn by CICADA, and let $K = K_0 + K_a$ be the total number of curves. Then, we can define a penalisation term as:
\begin{align} \label{eq:bezier_loss}
    L_b(x; \bar{x}) \doteq \sum_{k=1}^{K_0} \sum_{m=1}^3 \lambda_m \|\bar{x}_k^{(m)} - x_{k}^{(m)}\|_2^2,
\end{align}
where $\lambda_m>0$ are regularisation parameters weighing the importance of matching the original curves for every type of variable. This approach poses the advantage that we can tune those parameters depending on how much freedom we want to give the agent on adjusting the traces, as greater values of $\lambda_m$ will tighten the restrictions and smaller values will loosen them. The $\lambda_m$ parameters can even be chosen dependant on the traces, so as to loosen the restrictions on the parts of the drawing of which the user is less confident.

While this is a very good method for enforcing the user-drawn curves to be kept in place during the design process, it does not entirely enforce that the drawing stays unaltered. In fact, our agent would still be capable of drawing on top of the provided sketch (and often will) even to the point of completely obscuring it, if given a large enough number of curves.

\subsection{Geometric Fidelity}

In order to prevent the agent from drawing on top of the user-provided sketch (or rather from abusing its capacity to do so) previous attempts were made at using a square-distance penalisation approach \cite{ibarrola2022towards}. However, this type of pixel-by-pixel penalisation can often be too rigid, as identical objects in adjacent but non-overlapping positions are penalised as heavily as if they were on opposite corners of a canvas. Intuitively, this means we are implying that moving something a little matters the same that moving it a lot, which in our context is clearly undesirable. This effect -- which  is particularly pronounced in a line drawing context -- can stifle the gradients needed for further optimisation. Other metrics can account for distance more effectively, such as the Earth Mover's Distance, but their computation is often costly, and our heavily interactive use case requires real-time optimisation progress.

A simpler answer lies in the very same NN that builds up the CLIP encoding function, as it has been shown in style transfer \cite{ecker2015neural} applications that different layers of a Convolutional Neural Network (CNN) account for different geometric properties of an image. Thus, if we consider CLIP's encoding function as a composition of its inner layers $f = f^{(Q)} \circ \ldots \circ f^{(2)} \circ f^{(1)}$, we can define a new penaliser in term of the layer outputs. Namely, if we use $\cdot|_{\Omega^c}$ to denote a masked image where the color of the pixels inside $\Omega$ is replaced by the background color, then we can attest similarity using the function
\begin{align} \label{eqn:geo_loss}
    L_\Omega(x; \bar{x}) = \alpha\sum_{q\in\mathcal{Q}} \|f^{(q)}(h(x)|_{\Omega^c})-f^{(q)}(h(\bar{x})|_{\Omega^c})\|_2,
\end{align}
where $\alpha >0$ is a regularisation parameter, and $\mathcal{Q} \subset \mathbb{N}_Q$ is the set of indices of the layers used for computing image fidelity. This will effectively deal with the issues related to spatial arrangement, since this loss is sensitive to positioning, not just overlapping. 

\subsection{Optimisation}

Having established all of the main terms, we can formulate our problem as that of minimising the cost function
\begin{align}\nonumber
    L(x;\bar{t}, \bar{x}) \doteq L_s(x ; \bar{t}) + L_b(x ; \bar{x}) + L_\Omega(x ; \bar{x}),
\end{align}
with respect to $x$.

In order to do that, we can make use of a gradient descent method
\begin{align} \label{eq:grad_desc_L}
    x \leftarrow x - \eta \nabla_x L(x;\bar{t}, \bar{x}),
\end{align}
where $\eta > 0$ is the descent step. It is worth mentioning here that in practice, we can use more advance gradient descent methods, such as ADAM. The code for this optimisation and the methods that follow is available online.\footnote{\href{https://github.com/fibarrola/cicada}{https://github.com/fibarrola/cicada}}

A problem often found when minimizing non-convex cost functions such as $L$ using iterative methods is that they are prone to converge to local minima. Nevertheless, since we have established that we are looking for a set of diverse solutions rather than a global optimum, this actually plays in our favour. In fact, if we define quality in terms of a loss function, then any set of diverse high-quality solutions is a subset of the local minimisers of said function. On the other hand, variety in the drawings does not always entail diversity on the designs. In fact, since CLIP was trained in natural images, it is often the case that the method produces undesirable scribbles attempting to match some background, but that can be mitigated using initialisation and pruning techniques.

\subsection{Treebranch initialisation}\label{sec:treebranch_initialisation}

The standard CLIPDraw initialisation method is based on uniformly sampled random points, which makes sense when drawing from scratch, but poses an issue for design completion. The proposed gradient descent optimisation can only move points within a small neighborhood around the previous point locations, and since every small displacement needs to entail a decrease in the loss, traces tend to end up very close to their starting points.  Given that most sketches have a central focus point on a blank canvas, this means many of these ``stuck'' traces become background noise. This is a form of the vanishing gradient problem \cite{hanin2018neural} specific to optimising sparse vector drawings.

To avoid this issue, we propose a ``treebranch'' initialisation method for the traces added to the design, using a partition of the set of new traces $T = T_1 \cup T_2 \cup T_3$.
\begin{enumerate}
    \item Initialise the traces of $T_1$ such that the starting points are randomly selected from the set of all the points from the user provided sketch inside the drawing region $\Omega$. The other points of the trace should be randomly Initialised at a fixed distance from the previous ones, the same as in CLIPDraw.
    \item Initialise the traces of $T_2$ such that the starting points are randomly selected from the set of all the endpoints in $T_1$.
    \item Initialise the traces of $T_3$ such that the starting points are uniformly sampled from the drawing region.
\end{enumerate}

Using this initialisation, we make sure that the new traces are positioned in a way that they are more likely to become part of the focal point of the sketch -- the ``design'' -- than the background.

\subsection{Pruning}\label{sec:pruning}

Given that the chosen optimisation process is local, CICADA greatly benefits from initialising an oversized number of additional traces for drawing completion. The drawback of this approach is that we end up with quite a lot of unnecessary scribbles on the canvas, making pruning some of them highly desirable. However, distinguishing between a scribble and a trace critical to the the drawing is an ill-defined task, as there is no ground truth to decide which is which, but merely a subjective, intuitive notion.

One approach to attest this would be removing a set of paths and see how much the semantic loss $L_s(x, \bar{t}) \doteq -\langle g(\bar{t}), f \circ h(x)\rangle$ changes. So we could think of removing the set of $M$ traces that least increases $L_s$, but that would amount to testing $K_a!/(M!(K_a-M)!)$ combination. Instead, we can think of compromising and removing the $M$ paths whose individual removals least increase $L_s$, therefore needing only $K_a$ evaluations.

On the other hand, let us consider the example of designing a mug. In that case, removing a path representing a painting on top should not negatively affect the semantic loss, since the drawing should still be recognised as a mug. In addition to semantic loss, we shall consider the minimum distance between the guide-points $x$ on a path and those of the partial sketch as a notion of how related the traces are to the drawing. This will tend to disregard scribbles that are separated from the actual sketch.

Keeping both ideas in mind, we can prune out the $M$ traces with the lowest scores
\begin{align}
    s_p = L_s(\{x_k : k\neq p\}; \bar{t}) -\beta \min_{1\leq k\leq K_0} \{ \min_{i,j}\|\bar{x}_{k,j}-x_{p,i}\| \}
\end{align}

where $\beta>0$ is a weighting parameter. It is worth noting at this point, that while pruning can help us get a ``cleaner'' sketch, it may also delete traces that could otherwise serve for inspiration.

\section{Experiments and Results}

\begin{figure*}
    \centering
    \includegraphics[width=0.49\columnwidth]{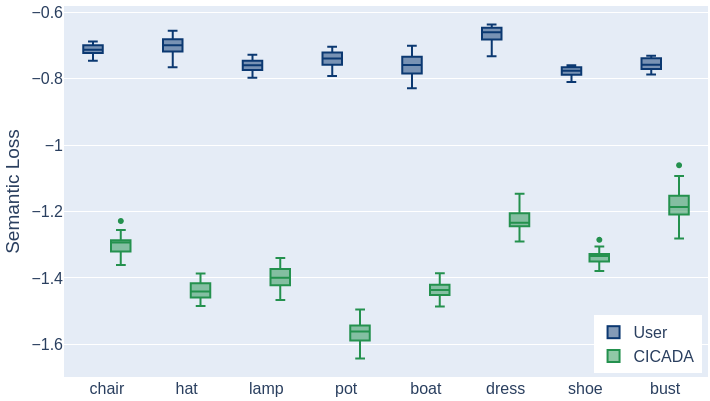}
    \includegraphics[width=0.49\columnwidth]{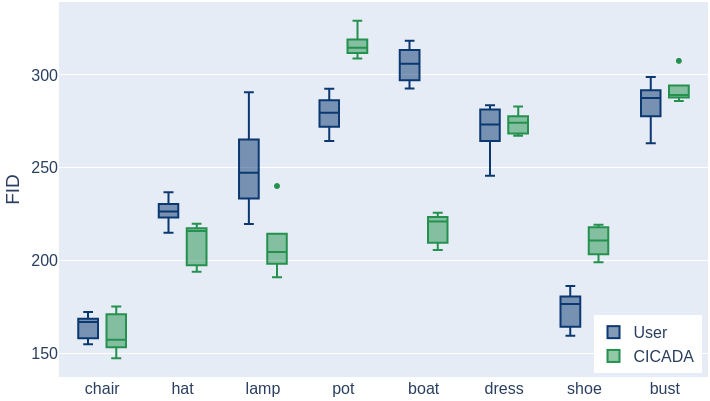}
    \caption{Left: Comparison of semantic losses of sketches as completed by users and CICADA. Right: FID between CICADA-completed sketches to user-completed sketches, and sketches completed by CICADA but with different parameters.}
    \label{fig:user_vs_cicada}
\end{figure*}

While CICADA is meant to be used in an iterative setting involving a human user and a computer, that requires a user interface, the design of which poses its own challenges (that go beyond the scope of this paper). Henceforth, in order to test CICADA, we will consider the problem of completing a partial sketch drawn by a human. Over those sketches, we can evaluate output creativity (in terms of quality and diversity, metrics suggested for collaborative creative AI in \cite{colton2020adapting}) and awareness (w.r.t. the user's input). However, we have established that this is not enough, given that there is a process in place, which from the model's perspective can be seen as iteratively adapting to changes made by the user to the prompt and the traces constituting the sketch. Hence, we evaluate for CICADA's adaptability to those two types of changes as a way to account for the design sketching task as a process.

Our evaluation process will heavily rely on measuring output diversity, to which there is no standard approach. In order to cope with this, we begin by establishing a way to do so in terms of the well known Shannon's Entropy \cite{shannon2001mathematical}.

\subsection{Measuring Output Diversity}

One of the most widely adopted methods for comparing sets of images is the Frech\'et Inception Distance (FID, \cite{heusel2017gans}). This is done by estimating the mean and covariance of the latent variables of the encoded values of the datapoints in an inception model (based on Imagenet, \cite{deng2009imagenet}), and has shown to correlate well with human judgements of quality. Nonetheless, there is not an established way of measuring diversity in our creativity context.

If we think of the output drawings as realisations of a random variable, as is done for computing the FID, then the output diversity can be interpreted as our uncertainty over such outputs, which is to say the entropy of the underlying probability distribution \cite{shannon2001mathematical}. 

Let us consider a set of images $A$, mapped to a multivariate normal distribution $f \sim \mathcal{N}(\mu, \Sigma)$ by the same methods used for computing FIDs. We could then assess the diversity of the set by computing the differential entropy $h$ of such distribution, defined as
\begin{align*}
    h(f) = -\mathbb{E}\log(f) = \frac{1}{2}\log\det(2\pi e \Sigma).
\end{align*}

Unfortunately, the dimension $D$ of the domain of $f$ is often too large for such determinant to be computed with a limited number $N$ of available samples. The estimated covariance matrix is computed as $\Sigma = B(I-\frac{1}{1-N}\mathbf{1}_{N\times N})^2B\T$, where $B \in \mathbb{R}^{D\times N}$ is the matrix concatenating the inception model's latent vectors for each of the $N$ samples in the dataset A, meaning that in practice $\text{range}(\Sigma) =  N-1$.

To overcome this, we define the Truncated Inception Entropy as
\begin{align}\label{eqn:pie}
    \text{TIE}(A) \doteq \frac{K}{2}\log(2\pi e) + \frac{1}{2}\sum_{k=1}^{K}\log\lambda_k,
\end{align}
where $\{\lambda_k\}$ are the non-zero eigenvalues of $\Sigma$, and $K=\min\{N-1, D\}$.  It can be shown that this definition matches the differential entropy when $\Sigma$ has full rank.

We shall henceforth use the TIE as an indicator of diversity over sets of images.

\subsection{Output Creativity}

In order for a set of generated sketches to be considered creative, we have established they have to attest for sufficient quality and diversity. To test for this, we asked 12 participants to complete a set of five randomly assigned sketches according to their respective text prompts, and compare them to sketches generated by CICADA.

Firstly, we compare the semantic loss $L_s$ of the user-completed sketches against those completed by CICADA. In Figure \ref{fig:user_vs_cicada}, we can see that the CICADA outputs greatly outperform the user generated sketches, but this is no great surprise since $L_s$ is a term of the loss our model is optimised to fit. CICADA is hence working correctly, yet this does not really speak to quality in accordance with human perception.

There is not a well established, automated way to get a sense of the quality of CICADA drawings with respect to those made by users. The most popular FID requires a point of reference that we do not have, but we can fabricate by changing CICADA's parameters. The second boxplot in Figure \ref{fig:user_vs_cicada} shows (in blue) the FID between sets of sketches completed by users and CICADA and (in green) FIDs between sets of sketches completed by CICADA and yet other CICADA-completed drawings, but using a higher pruning ratio. It can be seen that the resutls vary substantially by class, likely due to aspects of the different partial sketches used in our test (top row of Fig. \ref{fig:cicada_gen}), with 3/8 classes showing no significant differences between the User-CICADA and CICADA-CICADA FIDs, 3/8 showing greater user-CICADA FID, and the remaining two showing greater CICADA-CICADA FID.  The lack of a clear prevalence indicates similar quality in terms of what FID can measure.

\begin{figure}
    \centering
    \includegraphics[width=0.49\columnwidth]{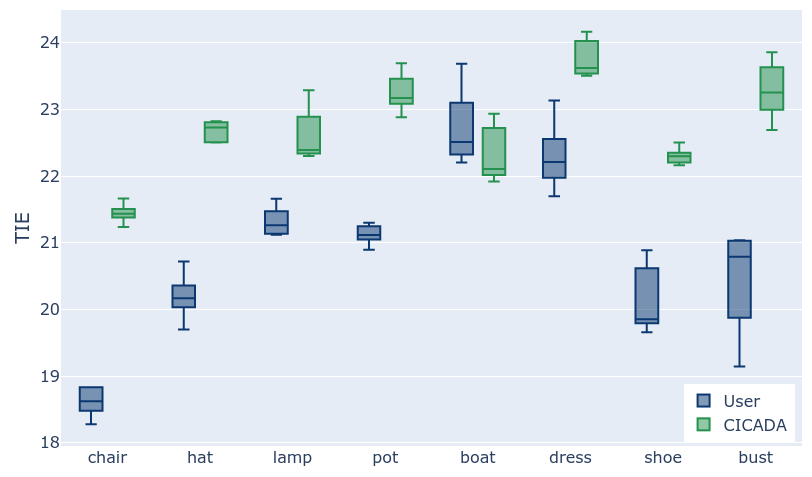}
    \caption{Diversity of sketches completed by users and CICADA as measured by TIE.}
    \label{fig:tie_user_vs_cicada}
\end{figure}

To account for diversity, sketch completion from both human users and CICADA were compared in terms of TIE. Figure \ref{fig:tie_user_vs_cicada} shows that there is a significantly ($p<0.05$) larger degree of diversity in favour of CICADA in all but one category, where they are rather similar.

In order to complement these automated results, we asked another set of eight participants to rate sketched completed by both CICADA and other users. We asked them to rate a random subset of six images per partial sketch from one to five on two scales: perceived quality as a representation of the prompt and novelty given the prompt. A number of the CICADA-generated images had some scribbles in the background even after the techniques described in Sections 2.5 and 2.5, and we pruned these errant strokes by hand: they are not relevant to the comparison and may cause confusion. In any practical use-case, this removal is expected to take place anyway. Some examples of the CICADA-generated images can be seen in Figure \ref{fig:cicada_gen}.

\begin{figure*}
    \centering
    \includegraphics[width = \textwidth]{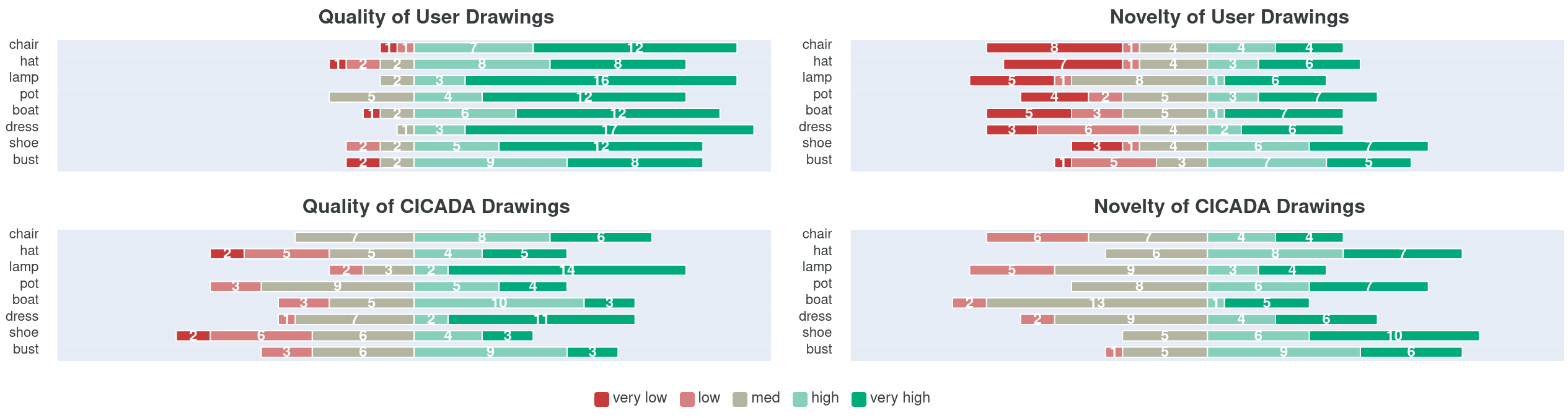}
    \caption{User ratings of sketches completed by other users (top) and by CICADA (bottom) in terms of quality (left) and novelty (right).}
    \label{fig:user_ratings}
\end{figure*}

The user ratings we obtained are depicted in Figure \ref{fig:user_ratings}. While the observed differences are moderate, the ratings do point to a tendency for better quality of the user-produced outputs over CICADA, while CICADA tends to produce a higher degree of novelty. This is overall aligned with the idea that CICADA can help the user explore new paths during the creative process, but also suggests the final decisions and finishing touches on the design remain a human's task.

\subsection{Awareness}

We want to assess CICADA's ``awareness'' -- how mindful it is of the user's sketch in progress. Too little would mean that a user's progress is disregarded (such as would happen by using CLIPDraw initalised with the user's strokes), whilst too much would mean leaving little room for creativity or re-interpretation (as in an inpainting framework where existing parts of the image are inviolate). In order to get a sense of this,  we can compare the CICADA outputs obtained by using very low, medium, or very high penalisation. A very low penalisation will make CICADA act like CLIPDraw (just with our custom initialisation and pruning), and a very high will act as if using CLIPDraw for inpainting (i.e. restricting it to the blank regions of the canvas. Additionally we would expect to see a decrease in diversity as the penalisation level is increased.

\begin{figure}
    \centering
    \includegraphics[width=0.49\columnwidth]{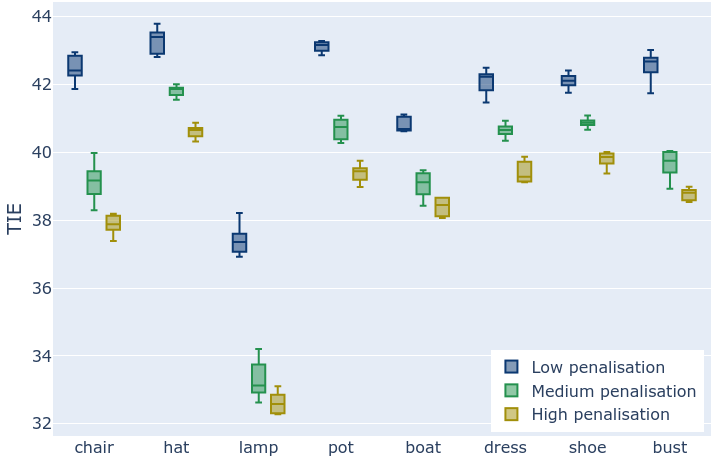}
    \caption{TIE of image sets generated using very low, medium or very high penalisation. Note that as penalisation increases, the TIE (our measure of generated image diversity) decreases, as the system is forced to hew more closely to the provided partial sketch.}
    \label{fig:tie_penalization}
\end{figure}

To corroborate this, we generated 30 drawings from each partial sketch, using three levels of penalisation. For a more robust comparison, we then built five sets of 20 randomly selected samples from those 30, and computed the TIEs to assess diversity over those sets. As can be seen in Figure \ref{fig:tie_penalization}, the TIE is lower for higher levels of penalisation, which is consistent with constraining some of aspects of the image to match more strongly those of the provided partial sketch.

\subsection{Adaptability}

Throughout the design process we expect users to regularly do two things: change their mind and redirecting the drawing. Here we test, using in-silico proxies, CICADA's capacity to effectively respond to these changes.

\subsubsection{Changing one's mind: CICADA's conceptual adaptability}

\begin{figure*}
    \centering
    \includegraphics[width=0.49\columnwidth]{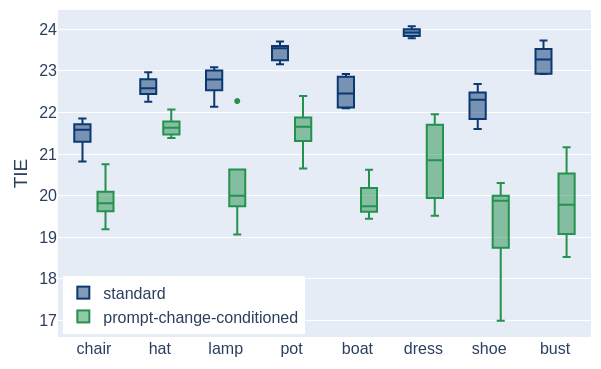}
    \includegraphics[width=0.49\columnwidth]{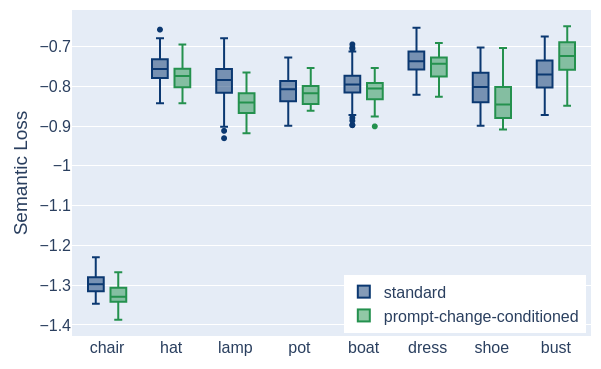}
    \caption{Left: TIE of the results as obtained with the standard CICADA algorithm vs those obtained when changing prompts. Right: Semantic loss $L_s$ of the same results. In both cases, the results are split into those obtained having started with a different prompt and then switching to a final prompt, and those obtained directly with the final prompts.}
    \label{fig:prompt_changes}
\end{figure*}

\begin{figure}
    \centering
    \includegraphics[width=0.49\columnwidth]{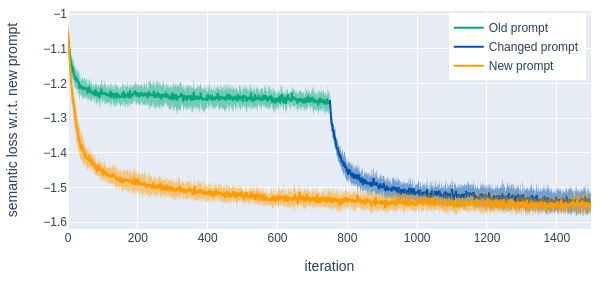}
    \caption{Semantic loss with respect to a prompt $t_1=\text{``A short blue chair''}$. The filled line indicates the mean and the shade the standard deviation (over 10 runs). In orange, the semantic loss w.r.t. $t_1$ for an image optimized to fit $t_1$. First in green and then in blue, the losses w.r.t. prompt $t_1$ for images first optimized to fit a different prompt $t_2=\text{``A tall red chair.''}$ (green), and then to fit prompt $t_1$ (blue). The final losses are not statistically significant ($p>0.05$).}
    \label{fig:prompt_change_loss}
\end{figure}

In our co-creative setting, users are expected -- as per literature in design cognition -- to continually rethink what they are doing, and hence change the direction of their design. This is likely to include making changes to the text prompt describing the task. After a prompt is changed it is necessary to pursue the new goal without entirely discarding the sketch progress so far. The co-creative model must therefore dynamically respond to the new prompt, adapting the previous developments towards the new goal.

For each of the eight partial sketches, we have first produced five samples using a prompt $t_2$. Using each of those five drawings as starting points, we changed the prompt to $t_1$ and let CICADA optimise the drawing to fit it, for 20 trials. On the other hand, we generated 30 sketches optimising to fit the prompt $t_1$ from the start. Fig \ref{fig:prompt_changes} illustrates the differences between the TIEs of the five sets of $t_2$-initialised results and five randomly sampled subsets of the 30 directly-optimised ones. It can be seen there is a significant TIE reduction across all classes, showing that the pre-optimisation for $t_2$ conditions the results, reducing their overall diversity and indicating that information from the original prompt is retained. Furthermore, the increase in variance among the changed-prompt FIDs suggests some sketches fit to match $t_2$ are more constraining than others when adapting to the new prompt $t_1$, in terms of the diversity of the results that can be obtained. We also show the semantic loss $L_s$ of the obtained results, showing comparable values, which is an indicator of the process not jeopardising the output quality.

Additionally, Fig. \ref{fig:prompt_change_loss} shows the changes in the loss on two processes: one where the user changes their mind about the prompt halfway, and another where CICADA starts with the final prompt from the start. The plot is loss against the new prompt only: the partial decrease in the (green) loss at the beginning has to do with the fact that the prompts contain some similarities (``a short blue chair'' vs ``a tall red chair''), chosen so as to mimic changes that might occur in a design process, where the designer is not likely to suddenly decide that they are designing something entirely different. When we switch prompts (blue), the loss immediately starts to decay and rapidly becomes effectively indistinguishable from a drawing directly optimised for that new prompt. Note we are displaying the results for ten independent runs, and there are still observable oscillations in the loss. A  $t-$test of the last 10 iterations showed no significant difference ($p>0.05$).

\subsubsection{Steering the drawing: CICADA's visual adaptability}

\begin{figure*}
    \centering
    \includegraphics[width=0.49\columnwidth]{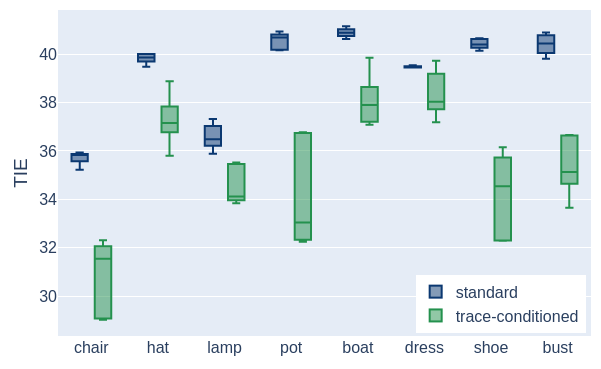}
    \includegraphics[width=0.49\columnwidth]{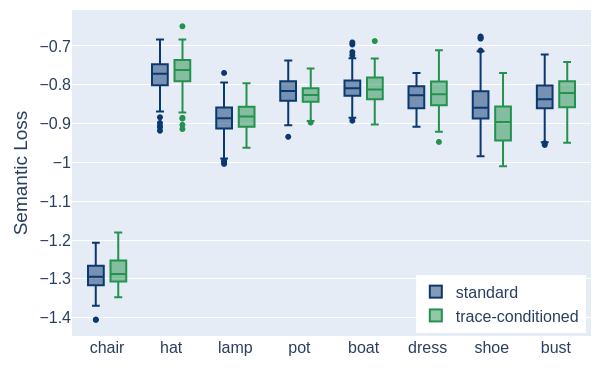}
    \caption{Left: TIE of obtained results. Right: Semantic loss $L_s$ of results. In both cases, the results are split into those obtained with and without being conditioned by a subset of fixed traces.}
    \label{fig:fixed_traces}
\end{figure*}

Collaboration is all about give and take, and in a co-creative drawing context a system will need to respond to partial acceptance and partial rejection of changes made by the system. When a user changes some strokes and discards others, CICADA needs to be able to produce new results that incorporate those changes. This parallels the ``conceptual adapability'' scenario above, in which the prompt is changed, but in this case CICADA needs to adapt to changes in the sketch. A co-creative drawing system will need to account for the influence of those changes in the outputs without sacrificing quality, which again we can measure based on changes in diversity (TIE) and loss between modified and unmodified sets.

For each of the eight partial sketches, we first produced 30 samples in the standard (i.e. uninterrupted generation) fashion, and randomly chose five of them. We heavily pruned the sketches, leaving just six strokes using the method described in Section \ref{sec:pruning}, to simulate a user's choice of traces to keep. With each of those sets as conditioning traces, we run CICADA's optimisation 20 times for each one, to obtain a set on which to compute the TIE. Finally, we calculated the TIEs for five random selections of 20 samples from the original 30 standard samples, comparing it to the the five trace-conditioned sets, as can be seen in Fig \ref{fig:fixed_traces}. To check for quality loss, we also computed the semantic losses (as defined in Equation \ref{eqn:semantic_loss}) with respect to the corresponding prompts.

There was a significant decrease in the TIE for the trace-conditioned sets across all eight categories, with no increase in the semantic loss (student's t-test on the per-category TIE and loss, $p <0.05$ as the threshold, Holm-Bonferroni \cite{aickin1996adjusting} correction applied). This means that the addition of traces is constraining the end result without jeopardising quality (in terms of semantic loss). As with the previous experiment there was a significant increase in the variance of the TIE diversity measure between runs, which points towards some choices of traces to fix being more restrictive than others.

\section{Discussion}

Upon assessing the experimental results there are some findings regarding CICADA's capabilities in terms of output creativity, awareness and adaptability. Firstly, given a partial sketch and an accompanying prompt, CICADA has been shown to be capable of producing creative outputs as accounted for in terms of quality and diversity. Our modified loss functions improve on previous vector-based synthesis by optimisation approaches to enable the integration of a partial sketch in a flexible but coherent way. This permits deployment to our co-creative drawing context in ways not facilitated by existing models.

CICADA outputs align well with the provided prompts, although as low-fidelity vector drawings they do not match the quality observed in current state-of-the-art raster image-based models such as DallE 2 \cite{ramesh2022hierarchical} or Imagen \cite{saharia2022photorealistic}. However, vector outputs compatible with the SVG framework permit native integration with user-authored drawings. Additionally, the fact that the CICADA drawings have a lower semantic loss than drawings produced by users suggests that optimiser capability is not a bottleneck, and  that performance may improve once loss functions that better align with user expectations are available.

CICADA outputs are also diverse, both as measured by our participants' ratings and our TIE diversity measure derived from FID. The capacity for diverse output is doubly critical in a co-creative context: firstly in order to support the many possible interpretations a user may have in mind when they provide a prompt, and secondly to facilitate adaptability as the user changes the drawing and/or the prompt. Very low output diversity would suggest significant conceptual brittleness, and would be unlikely to lead to high-quality output after user intervention.

It has also been observed that CICADA is aware or ``respectful'' of the user's contributions to the drawing process. This is demonstrated by the variations of TIE scores when changing the level of penalisation. The addition of penalisation terms make our model capable of  ``taking into account'' the current state of the design, modifying it without ever disregarding it entirely. This means that it has the potential to take an idea being elaborated by the user and steer it in a novel (yet still prompt-conditioned) direction, which may aid user creativity and prevent premature convergence (A phenomenon familiar to us ML researchers that turns out to occur as well in humans, where it is known as design fixation \cite{jansson1991design}). We hypothesise that these capabilities constitute an asset in augmenting creativity when deployed in a human-computer collaborative drawing context.

The capacity for cooperation is been further aided by CICADA's adaptability, both conceptual and visual.  Our model can accommodate a user changing their mind about some aspect of their drawing goal: should the user decide to change the prompt, the loss function is changed, and the model begins optimising for the new loss, initialised with the current canvas state. Similarly, users of a future CICADA-based interface will be able to modify (add, move, remove, rotate, scale, etc) curves at any point and the optimiser will re-initialise and then continue. Our experiments show that in both of these contexts (changing the sketch and changing the prompt) the existing sketch has an influence on the final sketch, meaning that the work done up to that point is not discarded, with no detriment on quality (at least as measured by the semantic loss).

Finally, it is worth mentioning that the design and development of a user interface that integrates CICADA is undergoing. A preliminary exploratory study suggests the system has capacity to facilitate co-creative drawing, with the majority of participants stating that it was a useful way to explore early ideas and visual concepts, although further experimentation and development is needed. Some subjects mentioned potential utility in specific design domains, such as character creation and interior design. Future evaluation of the complete system will explore its efficacy more rigorously, we mention our work to date solely in support of what our experiments here suggest: that the adaptations embedded in CICADA facilitate co-creative design.

\section{Conclusion}

We have operationalised co-creative drawing AI as an ML problem by drawing on literature in the cognitive science of design. We defined co-creative drawing not as the production of sketches to be used as design output, but the iterative negotiation of ideas through drawing. We then proposed CICADA, a vector-based, CLIP-guided, synthesis-by-optimisation method intended for the co-creative drawing context.  We show that CICADA is suitable for co-creative drawing in that it can 1) produce diverse and high-quality sketches, 2) adapt to changes in the prompt, and 3) adapt to changes in the drawing.  In evaluating our model we propose TIE, or Truncated Inception Entropy, as a measure of the diversity of a set of generated images.  We apply this diversity measure directly to the task of estimating CICADA's diversity when compared to humans, but also to the task of demosntrating that partial sketches provided by the user have an impact on CICADA's drawing (in that they reduce the diversity of what it can generate -- indicating that its outputs are conditioned on what was provided. We argue that these qualities make CICADA well suited to the co-creative drawing context we defined.

We are developing a user interface that will allow CICADA to be integrated into an interactive co-creative drawing software. Work also continues on other aspects of the model, where there is still much room for improvement, such as defining more effective, loss functions, faster methods for pruning, and mechanisms to focus on specific parts of the drawing.

\bibliographystyle{unsrt}

\end{document}